\documentclass[showpacs,preprintnumbers,prd,showkeys,aps]{revtex4}
\usepackage{eurosym}
\usepackage{amssymb}
\usepackage{amsfonts}
\usepackage{amsmath}
\usepackage{graphicx}
\usepackage{calrsfs}

\begin{document}

\title{GUP Assisted Hawking Radiation of Rotating Acoustic Black Holes }
\author{I. Sakalli}
\email{izzet.sakalli@emu.edu.tr}
\affiliation{Physics Department, Eastern Mediterranean University, Famagusta, Northern
Cyprus, Mersin 10, Turkey}
\author{K. Jusufi}
\email{kimet.jusufi@unite.edu.mk}
\affiliation{Physics Department, State University of Tetovo, Ilinden Street nn, 1200,
Macedonia}
\author{A. \"{O}vg\"{u}n}
\email{ali.ovgun@emu.edu.tr}
\affiliation{Physics Department, Eastern Mediterranean University, Famagusta, Northern
Cyprus, Mersin 10, Turkey}
\date{\today }

\begin{abstract}
Recent studies \cite{Jeff} provide compelling evidences that Hawking
radiation could be experimentally proven by using an analogue black hole. In
this paper, taking this situation into account we study the quantum
gravitational effects on the Hawking radiation of rotating acoustic black
holes. For this purpose, we consider the generalized uncertainty principle
(GUP) in the phenomenon of quantum tunneling. We firstly take the modified
commutation relations into account to compute the GUP modified Hawking
temperature when the massive scalar particles tunnel from this black hole.
Then, we find a remarkably instructive expression for the GUP entropy to derive the quantum gravity corrected Hawking
temperature of the rotating acoustic black hole.
\end{abstract}

\maketitle

\section{Introduction}

In a landmark agreement of general relativity and quantum field theory
(known as quantum gravity theory) Hawking \cite{Haw1} predicted that a black
hole (BH) could emit a blackbody like radiation which is the so-called
Hawking radiation (HR). In fact, HR asserts that any (not naked) BH, due to
the vacuum fluctuations, can evacuate the quantum particles generated from
the virtual particle pairs which obey the annihilation-creation mechanism of
the quantum field theory. It is hypothesized that in each HR process, one of
these particles which has negative energy falls into the BH to reduce its
mass, before being annihilated by its spouse (the one with positive energy).
Due to the vacua difference the non-absorbed particle can thus escape to
spatial infinity, and it could be detected by an observer as a HR particle.
The corresponding temperature of the HR is in striking agreement with the
first law of thermodynamics which comprises the entropy of the BH \cite{Bek1}%
. Unfortunately, the theoretical computations reveal that with today's
technology the HR of a cosmological BH (having mass of a few solar masses or
greater) is almost impossible. Because, such BHs have extremely a weak
radiation. However, using a different spacetime configuration one can get
arbitrarily strong HR. In particular, HR around a micro BH can tear it apart
relatively quickly \cite{Adam}.

Unruh \cite{un01} proposed a theoretical method which renders the HR
detection possible by stimulating an analogue BH (a quantum fluid originated
from Bose-Einstein condensation) in a laboratory environment. His argument
was based on the fact that a transition from subsonic flow to supersonic
flow is analogous to a BH event horizon. After then, numerous systems
(ultracold fermions, electromagnetic waveguides, light in a nonlinear
liquid, etc.) have been proposed for the analogue BHs (see \cite%
{Leonhardt,Nambu,Carbonaro,Iorio,Richartz} and references therein). For
example, Corley and Jacobson \cite{Corley} came up with a new idea for the
possible detection of the HR, which is about condensed matter BH laser.
According to their Gedanken experiment, when a BH possesses two horizons, HR
reflects between those horizons and by this way, in each round trip, an amplified radiation could disperse around the BH: natural increase in
the probability of HR detection.

Since Unruh's seminal papers \cite{un1}, it has been understood that in the
theory of supersonic acoustic flows the BH physics could be mimicked. The
obtained acoustic black holes (ABHs) govern the propagation of sound, which
depends algebraically on the flow velocity and density \cite{PLB1,PLB2}%
. Recently remarkable progress has been achieved in modeling the HR within
the laboratory \cite{Jeff}. So, these developments will not only help the
physicists to understand the most profound insights about the quantum
gravity, but they increase the popularity of the ABHs in the literature, as
well. The HR of the ABH was firstly studied by Li-Chun et al. \cite{Chun}
who employed two different methods in their computations: the reduced global
embedding approach and analytical continuation method of wave function
across the acoustic horizon. The main aim of this paper is to continue the
spirit of \cite{Chun} and give supplementary quantum gravity effects on the
HR of the ABH by focusing on the GUP \cite{GUP0,gup01}.

The remainder of the paper is organized as follows. In Sec. 2, we begin with
the spacetime metric for a rotating ABH and discuss some of its basic
features. We also show that the associated spacetime metric can be obtained
in a static metric structure by performing a simple dragging coordinate
transformation. In Sec. 3, we study how a massive chargeless scalar field
propagates under the effect of GUP in the background of the ABH. Writing out
the associated Klein-Gordon equation (KGE) with GUP, we show that KGE (within the framework of the Hamilton-Jacobi (HJ) method \cite{Vanz}) completely
separates with a suitable ansatz and yields GUP corrected Hawking
temperature. Section 4 is devoted to derivation of the quantum gravity
corrected (QGC) entropy, which we call it simply GUP entropy ($S_{GUP}$). In
Sec. 5, we explore the quantum gravity effects on the HR of ABH arising from
that of $S_{GUP}$. To this end, we perform the quantum tunneling
computations prescribed by Parikh and Wilczek \cite{PWT}. The paper ends
with Conclusion in Sec. 6.

\section{Rotating ABH}

The spacetime of the rotating ABH was introduced in \cite{PLB1,PLB2}.
Its line-element reads
\begin{equation}
ds^{2}=-Fdt^{2}+\frac{1}{G}dr^{2}-Hd\varphi dt+Kd\varphi ^{2},  \label{1}
\end{equation}%
where the metric functions are given by%
\begin{eqnarray}
F &=&\frac{1}{\lambda _{+}}\left( 1-\frac{\lambda _{-}(A^{2}+B^{2})}{%
c^{2}r^{2}}\right) ,\,\,\,\,  \label{2} \\
G &=&\frac{\lambda _{-}}{\lambda _{+}}\left( 1-\frac{\lambda _{-}A^{2}}{%
c^{2}r^{2}}\right) ,\,\,\,\,  \label{3} \\
H &=&\frac{2B}{c},  \label{4} \\
K &=&\frac{\lambda _{+}r^{2}}{\lambda _{-}}.  \label{5}
\end{eqnarray}

Here $A$, $B$ and $\beta $ are the real constants. $\lambda _{\pm }=1\pm
\beta $ and $c=\sqrt{\frac{dh}{d\rho }}$ denotes the speed of sound.
Throughout the paper, without loss of generality, we assume that the
constant $A$ has a positive definite value. The event horizon ($r_{h}$) of the
rotating ABH is conditional on $g^{rr}(r_{h})=G(r_{h})=0$. So that one gets 
\begin{equation}
r_{h}=\frac{\sqrt{\lambda _{-}}A}{c}.  \label{6}
\end{equation}

On the other hand, the condition $g_{tt}(r_{e})=0$ gives the radius of the
ergosphere as follows

\begin{equation}
r_{e}=\frac{\sqrt{\lambda _{-}}\sqrt{A^{2}+B^{2}}}{c}.  \label{7}
\end{equation}

Here we are interested in the frame-dragging effect (a BH drags spacetime
with it as it rotates) which is often termed dragging of inertial frames.
This effect produces a detectable gyroscopic precession called the
Lense-Thirring effect \cite{LTE}. When we use the dragging coordinate
transformation \cite{Jcap} $d\phi =d\varphi -\Omega dt,$ where $\Omega =%
\frac{H}{2K}$, metric \eqref{1} becomes 
\begin{equation}
ds^{2}=-Zdt^{2}+\frac{1}{G}dr^{2}+Kd\phi ^{2},  \label{8}
\end{equation}

where

\begin{equation}
Z=\frac{4KF+H^{2}}{4K}=\frac{c^{2}r^{2}-A^{2}\lambda _{-}}{c^{2}r^{2}\lambda
_{+}}.  \label{9}
\end{equation}

Hereby, the Hawking temperature of the rotating ABH can be computed \cite%
{Waldbook,rot3} as 
\begin{equation}
T_{H}=\frac{1}{4\pi }\sqrt{\frac{G}{Z}}\left. \frac{dZ}{dr}\right\vert
_{r=r_{h}}=\frac{\sqrt{\lambda _{-}}}{4\pi }\left. \frac{dZ}{dr}\right\vert
_{r=r_{h}}=\frac{c}{2\pi \lambda _{+}A}.  \label{10}
\end{equation}

\section{Effect of GUP on Quantum Tunneling of Scalar Particles}

Employing the modified commutation relations \cite{gup4}, it is shown in 
\cite{GUP0,GUP1,gup05,peng} that the KGE with GUP for a scalar field $\Psi $
takes the following form

\begin{equation}
-(i\hslash )^{2}\partial ^{t}\partial _{t}\Psi =\left[ (i\hslash
)^{2}\partial ^{i}\partial _{i}+m_{p}^{2}\right] \left[ 1-2\alpha
_{GUP}\left( (i\hslash )^{2}\partial ^{i}\partial _{i}+m_{p}^{2}\right) %
\right] \Psi ,  \label{11}
\end{equation}

where $\alpha _{GUP}$\ and $m_{p}$ are the GUP parameter and mass of the
scalar particle (phonon), respectively. The generalized KGE (11) can be
solved by using the semiclassical WKB approximation \cite{gup05}. We
therefore choose the following ansatz for the scalar field: 
\begin{equation}
\Psi (t,r,\phi )=\exp {\left( \frac{i}{\hbar }\mathcal{S}\left( t,r,\phi
\right) \right) }.  \label{12}
\end{equation}

where $\mathcal{S}( t,r,\phi)$ is the classically forbidden action for the tunneling. Inserting the above scalar field $\Psi $ into Eq. \eqref{11} for the
background \eqref{8}, we obtain the following expression (in leading order
of $\hbar )$ 
\begin{equation}
\frac{1}{Z}(\partial _{t}\mathcal{S})^{2}=G\,(\partial _{r}\mathcal{S})^{2}+%
\frac{1}{K}(\partial _{\phi }\mathcal{S})^{2}+m_{p}^{2}\left( 1-2\,\alpha
_{GUP}\,G(\partial _{r}\mathcal{S})^{2}-\frac{2\alpha _{GUP}}{K}(\partial
_{\phi }\mathcal{S})^{2}-2\alpha _{GUP}m_{p}^{2}\right) .  \label{13}
\end{equation}

Taking the symmetries of the metric \eqref{8} into account, one can choose
the following HJ ansatz for the action 
\begin{equation}
\mathcal{S}(t,r,\phi )=-E\,t+W(r)+j\phi +C,  \label{14}
\end{equation}%
where $C$ is a complex constant, $E$ is the energy, and $j$ denotes the
angular momentum of the particle. Substituting the action \eqref{14} into
Eq. \eqref{13}, one gets 
\begin{equation}
\frac{1}{Z}E^{2}=G\,(W^{\prime })^{2}+\frac{j^{2}}{K}+m_{p}^{2}\left(
1-2\alpha _{GUP}G(W^{\prime })^{2}-\frac{2\alpha _{GUP}}{K}j^{2}-2\alpha
_{GUP}m_{p}^{2}\right) .  \label{15}
\end{equation}

We focus only on the radial trajectories. Therefore, the radial part (by
ignoring the higher order terms of $\alpha _{GUP}$) results in the following
integral 
\begin{equation}
W(r)=\pm \int \frac{1}{\sqrt{\Delta (r)}}\frac{\sqrt{E^{2}-\frac{\Delta
\left( r\right) }{G}\left( \frac{j^{2}}{K}+m_{p}^{2}\right)
\left( 1-2m_{p}^{2}\alpha _{GUP}\right) }}{\sqrt{1-2m_{p}^{2}\alpha _{GUP}}}%
dr,  \label{16}
\end{equation}%
where 
\begin{equation}
\Delta (r)=ZG=\frac{\lambda _{-}}{\lambda _{+}^{2}}\frac{\left(
c^{2}r^{2}-A^{2}\lambda _{-}\right) ^{2}}{c^{4}r^{4}},  \label{17}
\end{equation}

which vanishes at the event horizon; $\Delta (r_{h})\rightarrow 0$. In order
to work out the integral \eqref{16}, we first expand the function $\Delta (r)
$ in Taylor's series near the horizon 
\begin{equation}
\Delta (r)\approx \Delta (r_{h})+\Delta ^{\prime }(r_{h})(r-r_{h})+\frac{1}{2%
}\Delta ^{\prime \prime }(r_{h})(r-r_{h})^{2}.  \label{18}
\end{equation}

Then, we evaluate the integral around the pole located at $r_{h}$ by
deforming the contour. The result is given by 
\begin{equation}
W(r_{h})=\pm \frac{i\pi E}{2\sqrt{\lambda _{-}}}\frac{\lambda _{+}r_{h}}{%
\sqrt{1-2m_{p}^{2}\alpha _{GUP}}}.  \label{19}
\end{equation}

The positive (negative) sign indicates the outgoing (ingoing) phonon. At
this point, we should note that the famous factor two problem in the above
expression which yields the wrong tunneling rate can be fixed with a
procedure described in \cite{Akhmedova1}. Another way to overcome this
problem is to set the probability of ingoing phonons to $100\%$. Namely, we
have 
\begin{equation}
P_{-}\simeq e^{-2ImW_{-}}=1,  \label{20}
\end{equation}%
which leads to 
\begin{equation}
Im\mathcal{S}_{-}=ImW_{-}+ImC=0.  \label{21}
\end{equation}

On the other hand, for the outgoing phonon we have 
\begin{equation}
Im\mathcal{S}_{+}=ImW_{+}+ImC.  \label{22}
\end{equation}

From Eq. \eqref{19} it is not difficult to see that $W_{+}=-W_{-}$. Hence,
one reads the tunneling probability of the outgoing phonons as follows 
\begin{equation}
P_{+}=e^{-2Im\mathcal{S}_{+}}\simeq e^{-4ImW_{+}}.  \label{23}
\end{equation}

Finally, using Eqs. \eqref{20} and \eqref{23} the tunneling rate of phonons
becomes 
\begin{equation}
\Gamma =\frac{P_{+}}{P_{-}}\simeq e^{(-4ImW_{+})}.  \label{24}
\end{equation}

Ultimately, we can find the GUP temperature ($T_{GUP}$) of the ABH by
comparing the latter result with the Boltzmann formula $\Gamma =e^{-\beta E}$%
, where $\beta $ is the inverse temperature \cite{rbook}. Thus, we have 
\begin{equation}
T_{GUP}=\frac{\sqrt{\lambda _{-}}}{2\pi \lambda _{+}}\frac{\sqrt{%
1-2m_{p}^{2}\alpha _{GUP}}}{r_{h}}=T_{H}\sqrt{1-2m_{p}^{2}\alpha _{GUP}}.
\label{25}
\end{equation}

As can be seen above, after terminating the GUP parameter i.e., $\alpha _{GUP}=0$, one can
recover the original Hawking temperature \eqref{10}.

\section{GUP Entropy}

In this section, we shall revisit the recent studies \cite{Pasos0,Pasos} to
derive the $S_{GUP}$ for a BH. In general, the GUP is defined by \cite%
{Vagenas} 
\begin{equation}
\Delta x\Delta p_{GUP}\geq \hbar \left( 1-\frac{y}{\hbar }\Delta p_{GUP}+%
\frac{y^{2}}{\hbar ^{2}}(\Delta p_{GUP})^{2}\right) ,  \label{26n}
\end{equation}

where $y=\alpha _{GUP}l_{p}$ in which $\alpha _{GUP}$ is a dimensionless
positive constant and $l_{p}=\sqrt{\frac{\hbar G}{c^{3}}}$ is the Planck
length. Equation (26) can be reorganized as

\begin{equation}
\Delta p_{GUP}\geq \frac{\hbar (\Delta x+y)}{2y^{2}}\left( 1-\sqrt{1-\frac{%
4y^{2}}{(\Delta x+y)^{2}}}\right) ,  \label{27n}
\end{equation}

In fact $l_{p}/\Delta x$ is infinitesimally small compared with unity.
Without loss our generality, using units $l_{p}=G=c=\hbar =k_{B}=1 $ and in
sequel expanding the above equation in Taylor series, we find out 
\begin{equation}
\Delta p_{GUP}\geq \frac{1}{\Delta x}\left[ 1-\frac{\alpha _{GUP}}{2\Delta x}%
+\frac{\alpha _{GUP}^{2}}{2(\Delta x)^{2}}+\cdots \right] .  \label{28n}
\end{equation}

As known from introductory quantum mechanics textbooks, in the absence of
the GUP effect ($\alpha _{GUP}=0$) we get the ordinary (Heisenberg) uncertainty
principle and its saturated form \cite{Pasos0,Pasos} as follows
\begin{equation}
\Delta x\Delta p\geq 1,  \label{29n}
\end{equation}

\begin{equation}
\Xi\Delta x\geq 1,  \label{30n}
\end{equation}

where $\Xi$ denotes the energy of a quantum-scale particle. Hence, getting
analogy between Eqs. (28) and (29), one can also derive the QGC version of
Eq. (30) as \cite{Pasos0} 
\begin{equation}
\Xi_{QGC}\geq \Xi\left[ 1-\frac{\alpha _{GUP}}{2(\Delta x)}+\frac{\alpha
_{GUP}^{2}}{2(\Delta x)^{2}}+\cdots \right] .  \label{31n}
\end{equation}%
The quantum tunneling rate for a quantum particle with $\Xi_{QGC}$ reads \cite%
{Pasos0} 
\begin{equation}
\Gamma \simeq \exp [-2Im\mathcal{I}]=\exp \left( -\Xi_{QGC}/T_{QGC}\right),
\label{32n}
\end{equation}%
where $T_{QGC}$ denotes the QGC temperature. Now, if we compare Eq. (32)
with the Boltzmann factor, we obtain 
\begin{equation}
T_{QGC}=T_{H}\left[ 1-\frac{\alpha _{GUP}}{2(\Delta x)}+\frac{\alpha
_{GUP}^{2}}{2(\Delta x)^{2}}+\cdots \right] ^{-1}.  \label{33n}
\end{equation}

Inspiring from the recent studies \cite{Pasos0,Pasos}, we can assign $\Delta
x$ to $A_{h}/\pi $. Thus, employing the first law of BH thermodynamics, one
can derive the GUP entropy as follows 
\begin{eqnarray}
S_{GUP} &=&\int \frac{\kappa dA_{h}}{8\pi T_{QGC}}=\int \frac{T_{H}dA_{h}}{%
4T_{QGC}}  \notag \\
&=&\int \frac{dA_{h}}{4}\left[ 1-\frac{\pi \alpha _{GUP}}{2A_{h}}+\frac{\pi
^{2}\alpha _{GUP}^{2}}{2A_{h}^{2}}+\cdots \right] ,  \notag \\
&=&\frac{A_{h}}{4}-\frac{\pi \alpha _{_{GUP}}}{8}\ln {\frac{A_{h}}{4}}-\frac{%
\pi ^{2}\alpha _{GUP}^{2}}{8A_{h}}+\cdots .  \label{34n}
\end{eqnarray}%
where $A_{h}$ is the perimeter length of the event horizon and $\kappa =2\pi
T_{H}$ is the surface gravity. In Eq. (34), the existence of $\alpha
_{_{GUP}}$ is brought correction terms to the BH entropy. Thus, whenever $%
\alpha _{_{GUP}}=0$ one can reproduce the well-known area law for the BH
mechanics: $\left. S_{GUP}\right\vert _{\alpha _{_{GUP}}=0}\rightarrow
S=A_{h}/4$. Meanwhile, the result obtained in Eq. (34) is in accordance with
the earlier works that take account of the influences of the loop quantum
gravity and string theory on the quantum corrected entropy (see, for
instance, \cite{Rovel,QGC1,QGC2,QGC3,QGC4} and references therein).

\section{Quantum Gravity Corrected Hawking Radiation of ABH}

The Painlev\'{e}-Gullstrand coordinate (PGC) \cite{Pain,Gull} system is one
of the coordinate transformations in general relativity that defines a
spacetime, which is regular at the horizon. The constant time surfaces in
the PGCs traverse the event horizon to reach the singularity. In fact, a
geometry described by the PGC can be seen as a flow whose current speed is
equal to the Newtonian escape velocity at each point. Furthermore, the PGC
time is the proper time of an observer who freely falls radially from rest 
\cite{Hamil,Kanai}. In this section, we shall define the PGC form of the ABH
and compute its HR via the HJ method \cite{Vanz}. Subsequently,
in the framework of the Parikh-Wilczek tunneling method (PWTM) \cite{PWT},
the QGC HR will be studied.

According to Eqs. (3) and (9), one can easily see that

\begin{equation}
G=\lambda _{-}Z.  \label{35n}
\end{equation}

Therefore metric (8) can be rewritten as

\begin{equation}
ds^{2}=-Zdt^{2}+\frac{1}{\lambda _{-}Z}dr^{2}+Kd\phi ^{2}.  \label{36n}
\end{equation}

After rescaling the radial coordinate to

\begin{equation}
r\rightarrow \sqrt{\lambda _{-}}\widetilde{r},  \label{37n}
\end{equation}

the latter metric (36) becomes

\begin{equation}
ds^{2}=-\widetilde{Z}dt^{2}+\frac{1}{\widetilde{Z}}d\widetilde{r}^{2}+%
\widetilde{K}d\phi ^{2},  \label{38nn}
\end{equation}

where

\begin{equation}
\widetilde{Z}=\frac{c^{2}\widetilde{r}^{2}-A^{2}}{c^{2}\widetilde{r}%
^{2}\lambda _{+}},  \label{39n}
\end{equation}

\begin{equation}
\widetilde{K}=\lambda _{+}\widetilde{r}^{2}.  \label{40nn}
\end{equation}

In metric (38) the event horizon corresponds to

\begin{equation}
\widetilde{r}_{h}=\frac{A}{c}.  \label{41nn}
\end{equation}

It is needless to say that the Hawking temperature (10) remains intact in this framework, as it should be. Now, one
can pass to the PGCs by applying the following transformation \cite{SakCTP}
to the metric (38)

\begin{equation}
d\widetilde{t}=dt+\frac{\sqrt{1-\widetilde{Z}}}{\widetilde{Z}}d\widetilde{r},
\label{42nn}
\end{equation}

where $\widetilde{t}$ is referred to the PGC time. In fact, as it can be
deduced from many introductory textbooks written on the general relativity, $%
\widetilde{t}$ is equivalent to the proper time in this coordinate system 
\cite{Rob}. Inserting the transformation (42) into metric (38), we get the
following line-element 
\begin{equation}
ds^{2}=-\widetilde{Z}d\widetilde{t}^{2}+2\sqrt{1-\widetilde{Z}}d\widetilde{t}%
d\widetilde{r}+d\widetilde{r}^{2}+\widetilde{K}d\phi ^{2}.  \label{43n}
\end{equation}

The relativistic HJ equation \cite{Angh}\ of the classical
action $\mathcal{I}$ is given by

\begin{equation}
g^{\mu \nu }\partial _{\mu }\mathcal{I}\partial _{\nu }\mathcal{I}+m^{2}=0,
\label{44nn}
\end{equation}

where $m$ is the mass of the phonon. For the metric (43), Eq. (44) results in

\begin{equation}
m^{2}-(\partial _{\widetilde{t}}\mathcal{I})^{2}+2\sqrt{1-\widetilde{Z}}%
(\partial _{\widetilde{t}}\mathcal{I})(\partial _{\widetilde{r}}\mathcal{I})+%
\widetilde{Z}(\partial _{\widetilde{r}}\mathcal{I})^{2}+\frac{1}{\widetilde{K%
}}(\partial _{\phi }\mathcal{I})^{2}=0.  \label{45n}
\end{equation}

Letting

\begin{equation}
\mathcal{I}=\mathcal{W}(\widetilde{r})+\mathcal{J}(\phi )-\mathcal{E}%
\widetilde{t},  \label{46n}
\end{equation}

and in sequel substituting the above ansatz in Eq. (45), we obtain

\begin{equation}
\mathcal{W}_{\left( \mathcal{\pm }\right) }=\int \frac{\mathcal{E}\sqrt{1-%
\widetilde{Z}}\pm \sqrt{\mathcal{E}^{2}-\widetilde{Z}\left( m^{2}+\frac{%
\left( \partial _{\phi }\mathcal{J}\right) ^{2}}{\widetilde{K}}\right) }}{%
\widetilde{Z}}d\widetilde{r}.  \label{47n}
\end{equation}

Thus, one can find that near the horizon Eq. (41) reduces to

\begin{equation}
\mathcal{W}_{\left( \mathcal{\pm }\right) }^{\mathcal{NH}}\mathcal{\approx E}%
\int \frac{1\pm 1}{\widetilde{Z}}d\widetilde{r}.  \label{48n}
\end{equation}

It is easy to see that $\mathcal{W}_{(-)}^{\mathcal{NH}}=0$ (i.e.,
probability of ingoing phonons $P_{-}=1$), which warrants $100\%$ absorption
of the ingoing phonons by the ABH. On the other hand, the integral of $%
\mathcal{W}_{(+)}^{\mathcal{NH}}$ has a pole at the event horizon. To
evaluate the associated integral, we use the Taylor series to expand the
metric function $\widetilde{Z}$ around the horizon $r_{h}$:

\begin{equation}
\widetilde{Z}\cong \widetilde{Z}^{\prime }(\widetilde{r}_{h})(\widetilde{r}-%
\widetilde{r}_{h})+\Game (\widetilde{r}-\widetilde{r}_{h})^{2}.  \label{49n}
\end{equation}

Then, after deforming the contour around the pole $r_{h}$, we obtain

\begin{eqnarray}
\mathcal{W}_{(+)}^{\mathcal{NH}} &\mathcal{\approx }&2\mathcal{E}\int \frac{d%
\widetilde{r}}{\widetilde{Z}},  \notag \\
&\mathcal{\approx }&2\mathcal{E}\int \frac{d\widetilde{r}}{\widetilde{Z}%
^{\prime }(\widetilde{r}_{h})(\widetilde{r}-\widetilde{r}_{h})},  \notag \\
&\mathcal{\approx }&i\pi \mathcal{E}\lambda _{+}\widetilde{r}_{h}.
\label{50n}
\end{eqnarray}

The above result yields the tunneling probability of the outgoing phonons as
follows

\begin{equation}
P_{+}=\exp (-2Im\mathcal{I)}=\exp \left[ -2Im\mathcal{W}_{(+)}^{\mathcal{NH}}%
\right] =\exp \left( -2\pi \mathcal{E}\lambda _{+}\widetilde{r}_{h}\right) ,
\label{51nn}
\end{equation}

which is also equal to the tunneling rate (since $P_{-}=1$):

\begin{equation}
\Gamma =\frac{P_{+}}{P_{-}}=\exp \left( -2\pi \mathcal{E}\lambda _{+}%
\widetilde{r}_{h}\right) .  \label{52nn}
\end{equation}

After recalling the Boltzmann formula $\left[ \Gamma =\exp (-\mathcal{E}/T)%
\right] $, we can read the horizon temperature of the ABH within the PGCs:

\begin{equation}
T_{PGC}=\frac{1}{2\pi \lambda _{+}\widetilde{r}_{h}}.  \label{53nn}
\end{equation}

The above result is in full agreement with the standard Hawking temperature
(10). Meanwhile, from the first law of thermodynamics $dE=T_{H}dS_{BH}$ one
can derive the thermodynamic energy as

\begin{equation}
E=\frac{2\ln \left( \widetilde{r}_{h}\right) }{\lambda _{+}}.  \label{54nn}
\end{equation}

We also want to extend our computations to the tunneling method which
considers the self-gravitation and back-reaction effects. To this end, we
employ the PWTM \cite{PWT}. In the PGCs,\ the radial null
geodesics of a test particle are defined by

\begin{equation}
\dot{\widetilde{r}}_{\left( \pm \right) }=\frac{d\widetilde{r}}{d\widetilde{t%
}}=\pm 1-\sqrt{1-\widetilde{Z}},  \label{55nn}
\end{equation}

where positive (negative) sign stands for the outgoing (ingoing) geodesics.
Hence, the near horizon radial outgoing null geodesics can be derived as

\begin{eqnarray}
\dot{\widetilde{r}}_{\left( +\right) } &\cong &1-\sqrt{1-\widetilde{Z}\left( 
\widetilde{r}_{h}\right) }+\frac{1}{2}\frac{\widetilde{Z}^{\prime }(%
\widetilde{r}_{h})}{\sqrt{1-\widetilde{Z}\left( \widetilde{r}_{h}\right) }}(%
\widetilde{r}-\widetilde{r}_{h})+\Game (\widetilde{r}-\widetilde{r}_{h})^{2},
\notag \\
&\approx &\kappa (\widetilde{r}-\widetilde{r}_{h}),  \label{56nn}
\end{eqnarray}

where $\kappa =\frac{\widetilde{Z}^{\prime }(\widetilde{r}_{h})}{2}$ is the
surface gravity \cite{Waldbook}. According to the PWTM, while the particle
tunnels the event horizon from in ($\widetilde{r}_{i}$) to out ($\widetilde{r%
}_{f}$), the BH is supposed to emit a circular shell of energy $\omega $
which is very small compared with the total (fixed) energy $E$ i.e., $\omega
\ll E$ \cite{PWT}. This event precipitates the energy of the ABH from $E$ to 
$E-\omega $. Having regard to this self-gravitational effect \cite{KWa}, the
imaginary part of the action becomes \cite{SakCTP,Zhan,Bane,SakIJTP}

\begin{align}
Im\mathcal{I}& =Im\int_{\widetilde{r}_{i}}^{\widetilde{r}_{f}}\int_{E}^{E-%
\omega }\frac{dH}{\dot{\widetilde{r}}_{\left( +\right) }}d\dot{\widetilde{r}}%
_{\left( +\right) },  \notag \\
& =-Im\int_{r_{in}}^{r_{out}}\int_{0}^{\omega }\frac{d\widetilde{\omega }}{%
\dot{\widetilde{r}}_{\left( +\right) }}d\dot{\widetilde{r}}_{\left( +\right)
},  \label{57nn}
\end{align}

where the Hamiltonian $H=E-\widetilde{\omega }$\ ($dH=-d\widetilde{\omega }$%
). Using Eq. (56), one can evaluate the integral (57) by deforming the
contour. Thus, one obtains

\begin{eqnarray}
Im\mathcal{I} &=&-\pi \int_{0}^{\omega }\frac{d\tilde{\omega}}{\kappa }=-%
\frac{1}{2}\int_{0}^{\omega }\frac{d\tilde{\omega}}{T_{H}},  \notag \\
&=&-\frac{1}{2}\int_{S(E)}^{S(E-\omega )}dS,  \notag \\
&=&-\frac{1}{2}\left[ S(E-\omega )-S(E)\right] ,  \notag \\
&=&-\frac{1}{2}\Delta S.  \label{58nn}
\end{eqnarray}

Herewith, the tunneling rate becomes \cite{PWT}

\begin{equation}
\Gamma \sim e^{-2Im\mathcal{I}}=e^{\Delta S}.  \label{59n}
\end{equation}

If we re-set the universal gravitational constant to $G(=l_{p}^{2})=1/8$
(keeping on $c=\hbar =1$), the entropy results in $S=\frac{A_{h}}{4G}%
=2A_{h}=4\pi \widetilde{r}_{h}$: twice of the perimeter length of the
horizon (see for example \cite{3bh,Akbar}). Then, ignoring the effect of the
higher order GUP effects, Eq. (34) becomes (see also \cite{QGC5})

\begin{equation}
S_{GUP}=2A_{h}-\frac{\pi \alpha _{GUP}}{16\sqrt{2}}\ln \left( 2A_{h}\right)
+\Game (\alpha _{GUP})^{2}.  \label{60nn}
\end{equation}

With the aid of Eq. (54) one can re-express the event horizon as $\widetilde{%
r}_{h}=\exp \left( \frac{E}{2}\lambda _{+}\right) $. Whence, we can obtain
the change in the $S_{GUP}$ as

\begin{eqnarray}
\Delta S_{GUP} &=&S_{GUP}(E-\omega )-S_{GUP}(E),  \notag \\
&=&4\pi \left[ \exp \left( \frac{E-\omega }{2}\lambda _{+}\right) -\exp
\left( \frac{E}{2}\lambda _{+}\right) \right] +\frac{\pi }{32\sqrt{2}}\alpha
_{GUP}\lambda _{+}\omega .  \label{61}
\end{eqnarray}

When we expand Eq. (61) in the Taylor series with respect to $\omega $ and
accordingly arrange the terms up to the leading order in $\omega $, we get

\begin{equation}
\Delta S_{GUP}\cong -\left( \frac{1}{T_{H}}-\frac{\pi }{32\sqrt{2}}\alpha
_{GUP}\lambda _{+}\right) \omega +O(\omega ^{2}).  \label{62}
\end{equation}

Using Eq. (59), we can define the QGC tunneling rate as

\begin{equation}
\Gamma ^{QGC}\sim e^{\Delta S_{GUP}}=e^{-\frac{\omega }{T_{H}^{QGC}}},
\label{63}
\end{equation}

which yields the QGC Hawking temperature:

\begin{eqnarray}
T_{H}^{QGC} &=&\left( \frac{1}{T_{H}}-\frac{\pi }{32\sqrt{2}}\alpha
_{GUP}\lambda _{+}\right) ^{-1},  \notag \\
&=&T_{H}\left( 1-\frac{\pi }{32\sqrt{2}}\alpha _{GUP}\lambda
_{+}T_{H}\right) ^{-1},  \notag \\
&=&T_{H}\left( 1-\frac{\alpha _{GUP}}{64\sqrt{2}\widetilde{r}_{h}}\right)
^{-1}.  \label{64n}
\end{eqnarray}

One can compare Eqs. (64) and (33) (with $\Delta x=\frac{A_{h}}{\pi G}=\frac{%
8A_{h}}{\pi }=16\widetilde{r}_{h}$, $\alpha _{GUP}\equiv \alpha
_{GUP}l_{p}=\frac{\alpha _{GUP}}{2\sqrt{2}}$, and taking cognizance of the
leading order of $\alpha _{GUP}$) to verify that both temperatures obtained
are exactly the same. Furthermore, it is clear from Eq. (64) that ignorance
of the back-reaction effects ($\alpha _{GUP}=0$) regains the standard
Hawking temperature (10).

\section{Conclusion}

The HR of the rotating ABH in (2+1) dimensional spacetime was thoroughly
investigated in \cite{Chun}. However, it appears that GUP effects on that HR
has not been thoroughly studied in the literature. In this paper, we have
filled this gap by employing the KGE with GUP (11) and the GUP entropy (34).
For simplicity, we have considered the rotating ABH in the dragging
coordinate system. Next, we have demonstrated that the KGE with GUP for a
massive scalar field propagating in the background of an ABH completely
separates with the HJ ansatz. Then, focusing on the quantum
tunneling formalism, we have managed to find the GUP modified Hawking
temperature (25) of the ABH. Utilizing the GUP entropy derived in Sec. (4),
we have also obtained the QGC Hawking temperature (64). Both temperatures
have the standard Hawking temperature limit when the GUP effect is
terminated ($\alpha _{GUP}=0$).

\end{document}